\documentstyle[fleqn,espcrc2]{article}

\newcommand{\AmS}{{\protect\the\textfont2
  A\kern-.1667em\lower.5ex\hbox{M}\kern-.125emS}}

\newcommand{\be}{\begin{equation}}
\newcommand{\ee}{\end{equation}}
\newcommand{\bea}{\begin{eqnarray}}
\newcommand{\eea}{\end{eqnarray}}

\def\gtap{\;\raisebox{-.5ex}{\rlap{$\sim$}} \raisebox{.5ex}{$>$}\;}

\newcommand{\ia}{a^{-1}}

\newcommand{\x}{X_n^{\alpha}}
\newcommand{\xtwo}{X_{n+2}^{\alpha}}

\title{Lattice Monte Carlo Data versus Perturbation Theory.}

\author{Chris R. Allton\address{Department of Physics,
                                University of Wales, Swansea,
                                Singleton Park,
                                Swansea SA2 8PP,
                                U.K.}
\thanks{Preprint SWAT/128}}

\begin{document}

\begin{abstract}
Differences between lattice Monte Carlo data and perturbation theory
(for example the lack of asymptotic scaling) are usually associated
with the `bad' behaviour of the bare lattice coupling $g_0$ due to the
effects of large (and unknown) higher order terms in $g_0$. In this
philosophy a new, renormalised coupling $g'$ is defined with the aim of
making the higher order coefficients of the perturbative series in $g'$
as small as possible.

In this paper an alternative scenario is discussed where lattice
artifacts are proposed as the cause of the disagreement between Monte
Carlo data and the $g_0$-perturbative series. We find that with the
addition of a lattice artifact term, the usual asymptotic scaling
expression in $g_0$ is in excellent agreement with Monte Carlo data.
Lattice data studied includes the string tension, the hadronic scale
$r_0$, the discrete beta function, $M_\rho$, $f_\pi$ and the 1P-1S splitting
in charmonium.
\end{abstract}

\maketitle

\section{Introduction}
\label{sec:intro}

A necessary condition for lattice predictions of QCD and other
asymptotically free theories to have physical
(continuum) relevance is that they reproduce weak coupling
perturbation theory (PT) in the limit of the bare coupling $g_0\rightarrow 0$.
This perturbative scaling behaviour (a.k.a. asymptotic scaling) has not
yet been observed for complicated theories like QCD when the
{\em bare} lattice coupling $g_0$ is used as the expansion parameter.

As a result of this disappointing disagreement, various workers
have proposed methods of improving the convergence of the
perturbation series by a re-expansion in terms of some new coupling $g'$
\cite{parisi,lm}.

This paper studies an alternative viewpoint in which the disagreement
stems from lattice artifacts \cite{cra}. In this talk, it is shown
that these terms can provide the mismatch between the
lattice Monte Carlo data and $g_0$-PT without resorting to the use of a
re-defined coupling $g'$.

The QCD quantities studied in this analysis are:
the string tension,$\sqrt\sigma$;
the hadronic scale, $r_0$ \cite{sommer};
$M_\rho$; $f_\pi$; the $1P-1S$ splitting in charmonium;
and the discrete beta function $\Delta\beta$.

The results discussed here are presented in greater detail in \cite{main}.


\section{Lattice Distorted-Perturbation Theory}
\label{sec:ldpt_theory}

Two-loop perturbation theory predicts the
running of the lattice spacing $a$ with coupling $g^2$ as follows,
\be
\ia(g^2) = \frac{\Lambda}{f_{PT}(g^2)}, \;\;\;\;\;\; \mbox{where}
\label{eq:ia_PT}
\ee
\[
f_{PT}(g^2) = e^{- \frac{1}{2 b_0 g^2} } \;\; (g^2)^{-b_1 \over {2 b_0^2}}
\]

Lattice calculations predict $a$ by calculating some dimensionful
quantity on the lattice, and comparing it with it's experimental value.
As is well known these lattice values do not follow the above
perturbative behaviour (when the bare coupling $g_0$ is used).
There are a number of possible causes of the
disagreement: quenching; finite volume effects; unphysically large
value of the quark mass; a real non-perturbative effect; the inclusion
of only a finite number or terms (i.e. two) in the PT expansion;
and lattice artifacts due to the finiteness of $a$.
For the reasons outlined in \cite{cra}, the first three
effects cannot give rise to the sizeable discrepancy between
lattice data and PT.
As far as true (i.e. continuum) non-perturbative effects are concerned,
the overwhelming expectation is that for cut-offs of
$\ia \gtap 2$ GeV these effects should be minimal.
Therefore the disagreement can only be due to either or both
of the last two possibilities.

\begin{table*}[hbt]
\setlength{\tabcolsep}{.5pc}
%
\begin{tabular*}{\textwidth}{@{}l@{\extracolsep{\fill}}lllllll}
\hline 
{\bf Fitting}&        & \multicolumn{5}{c}{\mbox{\boldmath$a^{-1}$} {\bf from}} & \\
          \cline{3-7}
{\bf Method}&	& \multicolumn{1}{c}{\mbox{\boldmath$\sqrt\sigma$}}
                     & \multicolumn{1}{c}{\mbox{\boldmath$r_0$}}
                                & \multicolumn{1}{c}{\mbox{\boldmath$M_\rho$}}
                                                & \multicolumn{1}{c}{\mbox{\boldmath$f_\pi$}}
                                                          & \multicolumn{1}{c}{\mbox{\boldmath$1P-1S$}}
                                                                    & \multicolumn{1}{c}{\mbox{\boldmath$\Delta\beta(\beta)$}} \\
%
\hline 
\mbox{\boldmath$g_0$}{\bf -PT}
&$\Lambda$ [MeV]      & 1.254(3) & 1.599(5)  & 1.63(1) & 1.48(2) & 1.45(5) & ---   \\
&$\chi^2/dof$         & 484      & 262       & 10      & 9       & 6       & 702 \\
%
\hline 
{\bf Leading-}
&$\x$                 & 0.204(2) & 0.150(2)  & 0.22(2) & 0.34(3) & 0.35(6) & 0.373(5) \\
{\bf Order}
&$\Lambda$ [MeV]      & 1.90(1)  & 1.958(9)  & 2.15(5) & 2.2(1)  & 2.5(3)  & --- \\
{\bf LDPT}
&$\chi^2/dof$         & 3        & 16        & 1.1     & 1.6     & 0.3     & 4.2 \\
%
\hline 
{\bf Next-to-}
&$\x$                 & 0.26(2)  & 0.29(1)   & ---     & ---     & ---     & 0.24(1) \\
{\bf Leading-}
&$\xtwo$              &-0.024(6) &-0.046(3)  & ---     & ---     & ---     & 0.050(5) \\
{\bf Order}
&$\Lambda$ [MeV]      & 1.96(2)  & 2.14(2)   & ---     & ---     & ---     & ---      \\
{\bf LDPT}
&$\chi^2/dof$         & 1.7      & 1.4       & ---     & ---     & ---     & 1.7      \\
%
\hline 
{\mbox{\boldmath$g_{\overline{MS}}$}{\bf-PT}}
&$\Lambda$ [MeV]      & 17.34(4) & 20.89(7)  & 21.4(2)  & 21.0(3)  & 19.3(7) & --- \\
&$\chi^2/dof$         & 160      & 47        & 1.3      & 2.5      & 1.5     & 78 \\
%
\hline 
{\mbox{\boldmath$g_E$}{\bf -PT}}
&$\Lambda$ [MeV]      & 4.81(1)  & 5.56(2)   & 5.77(4)  & 5.58(7) & 5.2(2)  & --- \\
&$\chi^2/dof$         & 52       & 15        & 3.6      & 1.4     & 0.3     & 19 \\
%
\hline 
\end{tabular*}
\end{table*}

In this section we study the effect of lattice artifacts.
These can be parametrised (to leading order) by modifying
eq.(\ref{eq:ia_PT}) as follows:
\be
\ia_L(g_0^2) = \frac{\Lambda}{f_{PT}(g_0^2)}
\times  \bigg[ 1 - \x \frac{g_0^\alpha f_{PT}^n(g_0^2)}
                       {           f_{PT}^n(1)    } \bigg],
\label{eq:ldpt_fit}
\ee
(with no implicit summation over $\alpha$ or $n$).

Note that the ${\cal O}(a^n)$ coefficient in Eq(\ref{eq:ldpt_fit})
has been normalised so that $\x$
is the fractional amount of the ${\cal O}(g_0^\alpha a^n)$ correction
at a standard value of $g_0=1$ (i.e.$\beta=6$).
For $M_\rho, f_\pi$ and the $1P-1S$ splitting, $\alpha=0 \;\;\&\;\; n=1$;
for $\sigma$ and $\Delta\beta$, $\alpha=0 \;\;\&\;\; n=2$; and
for $r_0$, $\alpha=2 \;\;\&\;\; n=2$.

Lattice Monte Carlo data taken from many different collaborations
are fit to eq.(\ref{eq:ldpt_fit})
(see \cite{main} for a list of these references).
This provides the values for $\Lambda$,
$\x$, and the $\chi^2$ as listed in the table.
Also shown in the table are the fits to (2-loop) $g_0$-PT.
(The fit in this case is eq.(\ref{eq:ia_PT}) with $g\equiv g_0$.)

We see that leading order ``lattice distorted-PT'' fits the data very well
compared to the $g_0$-PT case, with the $\chi^2/dof$ down by an order of
magnitude or more.

As a further check of the method, we include in the fit the next-to-leading
term in $a$.
However, due to the large statistical errors in some of the lattice
data we perform this fit only for $\sigma, r_0$ and $\Delta\beta$
where the statistical errors are very small.
The fitting function appropriate for these quantities is:
\be
\ia_L(g_0^2) = \frac{\Lambda}{f_{PT}(g_0^2)}
\times  \bigg[ 1 - \x   \frac{g_0^\alpha f_{PT}^n(g_0^2)}
                        {           f_{PT}^n(1)    }
\label{eq:ldpt_fit2}
\ee
\[
\;\;\;\;\;\;\;\;\;\;\;\;\;\;\;\;\;\;\;\;
\;\;\;\;\;\;\;\;\;\;\;\;\;\;\;\;\;\;\;\;
            - \xtwo \frac{g_0^\alpha f_{PT}^{n+2}(g_0^2)}
                        {           f_{PT}^{\;n+2}(1)  } \bigg].
\]

The results of these fits are displayed in the table.

Obviously in the limit of infinite statistical precision,
adequate fits to the lattice distorted-PT formula would only be
obtained if the ${\cal O}(a^n)$ terms were included to all orders.
The fact that it is necessary to go to next-to-leading order
for the $\sigma$, $r_0$ and $\Delta\beta$ data to obtain
a sensible $\chi^2/dof$ simply states that these data
have sufficiently small statistical errors to warrant this
order fit.

The rest of this section comments on the results of these fits.

It is clear that for $\sigma, r_0$ and
$\Delta\beta$ data, the agreement between the data and
lattice distorted-PT is remarkable considering the tiny
statistical errors in the lattice data.

Another important finding is that the values of $\Lambda$ for the
various quantities are all consistent with $\Lambda=2.15$ MeV within
around $1\sigma$ with the only exception being the string tension.
This slight discrepancy can easily be explained by the effects of
quenching, and the uncertainties in the experimental value of $\sigma$.
Taking $\Lambda = 2.15 \pm 10\%$ MeV as an overall average, and
converting to the ${\overline{MS}}$ scheme, we have
$\Lambda^{(N_f=0)}_{\overline{MS}} = 190 \pm 20 MeV$.
This compares well with other lattice determinations and therefore
supports the validity of this approach.

The typical values of $\x$ in the table are 20-40\%.
A study at $\beta=6.4$ \cite{abada2} found that
non-perturbative determinations of the renormalisation constant of
the local vector current vary by 10-20\%
depending on the matrix element used.
Since the spread in $Z^{Ren}_V$ has been interpreted as ${\cal O}(a)$
effects \cite{heatlie}, we can assume that ${\cal O}(a)$ effects of
around 20-40\% are reasonable at $\beta=6.0$.

The coefficients for the second order terms are an order
of magnitude less than the first order terms. This follows
our expectation that the expansion in $f_{PT}$ in eq.(\ref{eq:ldpt_fit2})
forms a convergent series.

One of the most exciting features of the lattice distorted-PT
approach is that it can reproduce the behaviour of $\Delta\beta$.
The interpretation of the well-known discrepancy between $g_0$-PT and
Monte Carlo $\Delta\beta$ data has been problematic in the past.
For example, in \cite{taro}, a fit was attempted to their $\Delta\beta$ data
using a coupling with two free parameters.
A good fit was obtained only for an unphysical value of one of these
parameters, leaving the explanation of the discrepancy open.
The lattice distorted-PT approach solves this problem.

Finally, as far as the fit to $M_\rho, f_\pi$ and the $1P-1S$
splitting are concerned, the errors in the lattice data are
large enough to allow many functional forms. Thus these
data do not constrain the lattice distorted-PT fit
(or fits from other schemes).

\section{Fits Using a Renormalised Perturbation Theory}
\label{sec:g'}

In this section we fit the Monte Carlo data for $a^{-1}$ to the
following functional form:
\be
\ia_L(g_0^2) = \frac{\Lambda}{f_{PT}((g')^2)},
\label{eq:gprime_fit}
\ee
where $g'$ is some ``renormalised'' coupling which is in turn a function
of the bare coupling $g_0$.
Note that in this philosophy, the failure of asymptotic (i.e. perturbative)
scaling is explained by higher order terms in perturbation theory and
lattice artifacts are assumed to be negligible.


We studied two definitions of $g'$:
(i) The ${\overline{MS}}$-like coupling \cite{aida},
\[
\frac{1}{g_{\overline{MS}}^2(\pi/a)} = \frac{1}{g_0^2} \; <\frac{1}{3} Tr U_{plaq}>_{MC} + 0.025.
\]
and the scheme of \cite{parisi} based on the plaquette,
\[
\frac{1}{g_E^2} = \frac{1/3}{1 - <\frac{1}{3} Tr U_{plaq}>_{MC}}.
\]

(Note in the full paper, alternative
definitions of $g'$ are studied as well as those above \cite{main}.)

The results of fits using these definitions of $g'$ in the fitting function
Eq.(\ref{eq:gprime_fit}) are displayed in the table in
the rows headed $g_{\overline{MS}}$ and $g_E$.

As can be seen they do not reproduce the Monte Carlo results nearly as
well as the fits from the lattice distorted-PT.

\section{Conclusions}

This talk studies the question of why (dimensionful) lattice Monte
Carlo quantities do not follow the predictions of 2-loop perturbation
theory in the bare coupling. The conventional answer to this problem is
that higher order terms in $g_0^2$ spoil the behaviour of perturbation
theory, and that therefore an improved coupling is required. An
alternative approach is presented here where the effects of
${\cal O}(a)$ are shown to be able to provide the mismatch.
The quality of the fits using this latter approach, and various
arguments outlined in Sec.\ref{sec:ldpt_theory} support this
philosophy.
Further studies are required to unambiguously confirm this issue.



\end{document}